\documentclass[12pt]{JHEP3}

\preprint{  }

\usepackage{epsfig,multicol}
\usepackage{amsmath}
\usepackage{graphics}
\usepackage{graphicx}
\usepackage{dcolumn}
\usepackage{amssymb}
\usepackage{amsthm}
\usepackage{amsfonts}
\usepackage{subfigure}
\usepackage{setspace}

\title{Ehrenfest scheme for complex thermodynamic systems in full phase space}
\author{Zixu Zhao and Jiliang Jing\footnote{Corresponding author. Email:
jljing@hunnu.edu.cn}
\\ Department of Physics, and Key Laboratory of Low Dimensional Quantum Structures and Quantum Control of Ministry of Education, Hunan Normal University, Changsha, Hunan 410081, P. R. China}

\abstract{
For a thermodynamic system with multiple pairs of intensive/extensive variables and the thermodynamical coefficients attain finite or infinite values on the phase boundary, we obtain the two classes of Ehrenfest equations in the full phase space, and find that the rank of the matrix for these equations can tell us the dimensions of the phase boundary. We also apply this treatment to the RN-AdS black hole.}

\begin{document}

\section{Introduction}

The study of phase transitions has been a very active topic for a long time. For a usual thermodynamic system with two pairs of intensive/extensive variables, the first order and second order phase transitions can be described by Clausius-Clapeyron and Ehrenfest schemes, respectively. For another kind of phase transition, named the glass phase transition \cite{J}, the Ehrenfest relations along the glass transition line were used to find out whether there exists a thermodynamic description of the glass phase \cite{Nieuwenhuizen}.

The thermodynamic behavior of black holes have been studied extensively \cite{Davies,Hawking2} because black holes are considered as thermodynamic objects with entropy \cite{Bekenstein,Hawking}. In particular, several attempts have been made to describe the phase transition  \cite{Davies1,Hawking1,Carlip,Carter,Doneva} in black hole physics, and all these works are based on the fact that the phase transition in black hole thermodynamics is associated with the divergence of the specific heat. On the other hand, Gibbs \cite{Gibbs} found that the analysis of systems in thermodynamic
equilibrium can be considerably facilitated with the help of graphical and geometrical methods, which has the advantage of geometrically representing thermodynamic properties of simple fluids by means of
surfaces in ``Gibbs space" \cite{Tisza}. Then, Weinhold proposed an energy metric of equilibrium thermodynamics \cite{Weinhold,Weinhold2} and Ruppeiner introduced an entropy metric \cite{Ruppeiner,Ruppeiner2}, in which curvature singularities are related to phase transition points. Later, Quevedo showed the possibilities of studying thermodynamic systems in terms of geometric objects in an invariant manner \cite{Quevedo}.

The extension of Ehrenfest scheme to black hole thermodynamics seems to be quite natural since black holes behave as ordinary thermodynamic system in many respects. Such an attempt has been triggered recently \cite{Banerjee,Banerjee2,Banerjee3,Banerjee4,Banerjee5,Lala}.  Noting that the thermodynamical coefficients attain infinite values on the phase boundary in black holes, the authors in Refs. \cite{Banerjee,Banerjee2,Banerjee3,Banerjee4,Banerjee5,Lala} just took the usual Ehrenfest scheme as an approximation. The first application of the Ehrenfest¡¯s scheme to determine the phase transition for RN-AdS black hole was commenced in Ref. \cite{Banerjee3}, in which the numerical analysis was carried out to check the validity of the Ehrenfest equations approach to the critical point.
Recently, there has great interest to include the variation of the cosmological constant in the first law of black hole thermodynamics \cite{Caldarelli,Kastor,Dolan,Dolan2,Dolan3,C,L}. Kastor \cite{Kastor} showed that, by interpreting the cosmological constant as a pressure and its thermodynamically conjugate variable as a volume, one can get a complete form of the first law.
Especially, since Kubiz$\check{n}\acute{a}$k $\emph{et al}.$ \cite{KM} studied $P-V$ criticality of charged AdS black holes, much attention has been focused on the topic of applying Ehrenfest scheme to black holes very recently \cite{Mo,Mo2,Mo3,Mo4,Mo5}.
With respect to the RN-AdS black hole, it should be pointed that there are three pairs of intensive/extensive variables, i.e., temperature/entropy $(T,S)$, pressure/volume $(P,V)$ and electric potential/charge $(\Phi,Q)$ \cite{KM}. Selecting different subsystems in phase space, one could observe that there are different results by using the usual Ehrenfest scheme.

However, it should be noted that {\em{the usual Ehrenfest equations are only valid to analyze two pairs of intensive/extensive variables system and all thermodynamical coefficients attain finite values on the phase boundary}}. Thus, one can not directly apply the usual Ehrenfest equations to the RN-AdS black hole because this system has {\em{three pairs of intensive/extensive variables}} and  its thermodynamical coefficients attain {\em{infinite values on the phase boundary}}. How to apply the Ehrenfest scheme to complex thermodynamic systems (such as the RN-AdS black hole) is an important open question. In this paper, we try to address this question for a thermodynamic system with multiple pairs of intensive/extensive variables and the thermodynamical coefficients attain finite or infinite values on the phase boundary.

The organization of the work is as follows. In sec 2 we present Ehrenfest scheme in the full phase space. In sec 3 we apply Ehrenfest scheme to the RN-AdS black hole. We will conclude in sec 4 of our main results. We also discuss the situation of a thermodynamic system with $m$ ($m>3$) pairs of intensive/extensive variables in the appendix.

\section{Ehrenfest scheme in the full phase space}

In this section, we consider a thermodynamic system which has $(T,S)$, $(P,V)$ and a new pair of intensive/extensive variables $(X,Y)$. In order to find the Ehrenfest equations for this thermodynamic system, we express $dV$, $dS$ and $dY$ as
\begin{eqnarray} \label{dvdsdy}
&&dV=-V\kappa_{T,X}d P+V\alpha d T+V\gamma_{1}d X, \nonumber \\
&&dS=-V\alpha d P+\frac{ C_{P,X}}{T}d T+Y\gamma_{2}d X, \nonumber \\
&&dY=-V\gamma_{1}d P+Y\gamma_{2}d T+Y\gamma_{3}d X,
\end{eqnarray}
with
\begin{eqnarray}\label{coeff}
&&C_{P,X}=T\left(\frac{\partial S}{\partial T}\right)_{P,X}, \nonumber \\
&&\kappa_{T,X}=-\frac{1}{V}\left(\frac{\partial V}{\partial P}\right)_{T,X}, \nonumber \\
&&\alpha=\frac{1}{V}\left(\frac{\partial V}{\partial T}\right)_{P,X}, \nonumber\\
&&\gamma_{1}=\frac{1}{V}\left(\frac{\partial V}{\partial X}\right)_{P,T}, \nonumber \\
&&\gamma_{2}=\frac{1}{Y}\left(\frac{\partial Y}{\partial T}\right)_{P,X}, \nonumber \\
&&\gamma_{3}=\frac{1}{Y}\left(\frac{\partial Y}{\partial X}\right)_{P,T},
\end{eqnarray}
where we used the Maxwell relations $\left(\frac{\partial S}{\partial P}\right)_{X,T}=-\left(\frac{\partial V}{\partial T}\right)_{P,X}$, $\left(\frac{\partial S}{\partial X}\right)_{T,P}=\left(\frac{\partial Y}{\partial T}\right)_{P,X}$ and $\left(\frac{\partial V}{\partial X}\right)_{T,P}=-\left(\frac{\partial Y}{\partial P}\right)_{X,T}$. These thermodynamical coefficients exist the discontinuity on the phase boundary for the second order phase transition. On the phase boundary of second order phase transition, we have
\begin{eqnarray}\label{v1v2}
&&V^{(1)}=V^{(2)},~~~~~S^{(1)}=S^{(2)},~~~~~Y^{(1)}=Y^{(2)}, \nonumber \\
&&dV^{(1)}=dV^{(2)},~~dS^{(1)}=dS^{(2)},~~dY^{(1)}=dY^{(2)},
\end{eqnarray}
here and hereafter the superscript (1) and (2) denote phase 1 and phase 2, respectively.

We now face two cases that the coefficients (\ref{coeff}) attain finite or infinite values on the phase boundary. If the coefficients have finite values on the phase boundary, using Eqs. (\ref{dvdsdy}) and (\ref{v1v2}), we find the first class of Ehrenfest equations
\begin{equation}
\left(
                  \begin{array}{ccc}
1& -\frac{\Delta \alpha}{\Delta \kappa_{T,X}}  & -\frac{\Delta \gamma_{1}}{\Delta \kappa_{T,X}} \\
                   -\frac{TV\Delta \alpha}{\Delta C_{P,X}}  & 1 & \frac{TY \Delta \gamma_{2}}{\Delta C_{P,X}}\\
                  -\frac{V\Delta \gamma_{1}}{Y \Delta \gamma_{3}} & \frac{\Delta \gamma_{2}}{\Delta \gamma_{3}} & 1 \\
                  \end{array}
                \right)_{c}
\left(
\begin{array}{c}
dP \\
dT \\
dX \\
\end{array}
\right)=Z_{1c} x=0,\label{eeq}
\end{equation}
where $Z_{1c}$ and $x$ correspond to front matrix and subsequent matrix, here and hereafter the subscript $``c"$ denotes the critical values, and $\Delta \alpha=\alpha^{(1)}-\alpha^{(2)}$ and so on.

However, if these coefficients take infinite values on the phase boundary, near the boundary of phase $1$ or $2$, we have
\begin{equation}
\left(
                  \begin{array}{ccc}
1 & \frac{-\alpha}{\kappa_{T,X}}  & \frac{ -\gamma_{1}}{ \kappa_{T,X}} \\
                   -\frac{TV\alpha}{C_{P,X}}  & 1 & \frac{TY \gamma_{2}}{C_{P,X}}\\
                  -\frac{V\gamma_{1}}{ Y\gamma_{3}} & \frac{\gamma_{2}}{\gamma_{3}} & 1 \\
                  \end{array}
                \right)
\left(
\begin{array}{c}
dP \\
dT \\
dX \\
\end{array}
\right)=Z_2 x=\left(
\begin{array}{c}
\frac{-dV}{V\kappa_{T,X}} \\
\frac{T dS}{C_{P,X}} \\
\frac{dY}{Y\gamma_{3}} \\
\end{array}
\right).\label{eeq2}
\end{equation}
Then, on the phase boundary, we obtain
\begin{equation}
\left(
                  \begin{array}{ccc}
1 & -\frac{\alpha}{\kappa_{T,X}}  & -\frac{ \gamma_{1}}{ \kappa_{T,X}} \\
                   -\frac{TV\alpha}{C_{P,X}}  & 1 & \frac{TY \gamma_{2}}{C_{P,X}}\\
                  -\frac{V\gamma_{1}}{ Y\gamma_{3}} & \frac{\gamma_{2}}{\gamma_{3}} & 1 \\
                  \end{array}
                \right)_{c}
\left(
\begin{array}{c}
dP \\
dT \\
dX \\
\end{array}
\right)=Z_{2c} x=0,\label{eeq3}
\end{equation}
which are the second class of Ehrenfest equations.

Both the equations  (\ref{eeq}) and (\ref{eeq3}) have nontrivial solutions if $|Z_{ic}|=0$ ($i=1,2$), which means that the rank of matrix $Z_{ic}$ must be $r(Z_{ic})=1$ or $r(Z_{ic})=2$. It is of great interest to note that the second order phase transition will take place on a curved surface in the full phase space for $r(Z_{ic})=1$, or it will occur on a line for $r(Z_{ic})=2$.

\section{Applying to the RN-AdS black hole}

As an example, we now  consider the RN-AdS black hole. The temperature, entropy and electric potential are given by
$
T=\left(1+{3r_+^2}/{l^2}-{Q^2}/{r_+^2}\right)/{4\pi r_+}, S=\pi r_+^2$ and $\Phi={Q}/{r_+},\label{TSPhi}
$
where $l$ is the AdS radius and $r_+$ is the radius of the event horizon.
By using the suggestion that
$
P={3}/{8\pi l^2}$ and $ V={4\pi r_+^3}/{3},
$
the first law of the black hole thermodynamics \cite{KM} and the corresponding Smarr relation \cite{Kastor} can be expressed as
\begin{eqnarray}
&&dM=TdS+\Phi dQ+VdP,\nonumber \\
&&M=2(TS-PV)+\Phi Q.
\end{eqnarray}
We can easily show that the first order phase transition can be excluded in this system.
Taking $X=\Phi$ and $Y=Q$, we get
\begin{eqnarray}
&&C_{P,\Phi}=T\left(\frac{\partial S}{\partial T}\right)_{P,\Phi}=\frac{2S(8PS^2+S-\pi Q^2)}{8PS^2-S+\pi Q^2}, \nonumber \\
&&\kappa_{T,\Phi}=-\frac{1}{V}\left(\frac{\partial V}{\partial P}\right)_{T,\Phi}=\frac{24S^2}{8PS^2-S+\pi Q^2}, \nonumber \\
&&\alpha=\frac{1}{V}\left(\frac{\partial V}{\partial T}\right)_{P,\Phi}=\frac{12\sqrt{\pi}S^{3/2}}{8PS^2-S+\pi Q^2}, \nonumber \\
&&\gamma_{1}=\frac{1}{V}\left(\frac{\partial V}{\partial \Phi}\right)_{P,T}=\frac{6Q\sqrt{\pi S}}{8PS^2-S+\pi Q^2}, \nonumber \\
&&\gamma_{2}=\frac{1}{Q}\left(\frac{\partial Q}{\partial T}\right)_{P,\Phi}=\frac{4\sqrt{\pi} S^{3/2}}{8PS^2-S+\pi Q^2}, \nonumber \\
&&\gamma_{3}=\frac{1}{Q}\left(\frac{\partial Q}{\partial \Phi}\right)_{P,T}=\frac{\sqrt{S}}{Q\sqrt{\pi}}\frac{8PS^2-S+3\pi Q^2}{8PS^2-S+\pi Q^2},\nonumber \\
\end{eqnarray}
which show us that all coefficients become infinity when $8PS^2-S+\pi Q^2=0$. Therefore, we can use the second class of Ehrenfest equations (\ref{eeq3}) and  find $|Z_{2c}|=0$ and $r(Z_{2c})=1$. Therefore, the RN-AdS black hole undergoes a second order phase transition on a phase transition surface described by $8PS^2-S+\pi Q^2=0$.

\section{Conclusions and discussions}

It is well known that the usual Ehrenfest equations are only valid to analyze two pairs of intensive/extensive variables system and all thermodynamical coefficients attain finite values on the phase boundary. In this paper, we investigated a thermodynamic system with $(T,S)$, $(P,V)$ and a new pair of intensive/extensive variables $(X,Y)$ and the thermodynamical coefficients attain finite or infinite values on the phase boundary, and obtained the two classes of Ehrenfest equations in the full phase space. The first class of  Ehrenfest equations (\ref{eeq}) is valid to the case of these coefficients taking finite values on the phase boundary, which will reduce to the usual Ehrenfest equations if the system has no the new variables $(X,Y)$, and the second class of Ehrenfest equations (\ref{eeq3}) is valid to the case of these coefficients attaining infinite values on the phase boundary. It is of great interest to note that the rank of the matrix for Eq. (\ref{eeq}) or (\ref{eeq3}) can tell us that the second order phase transition will take place on a surface or on a line. By applying the Ehrenfest equations to the RN-AdS black hole, we showed that the black hole undergoes a second order phase transition on the phase transition surface described by $8PS^2-S+\pi Q^2=0$. It is natural to generalize this treatment to a thermodynamic system with $m$ ($m>3$) pairs of intensive/extensive variables (please see the appendix).

\section{Appendix}

With respect to a thermodynamic system with $m$ ($m>3$) pairs of intensive/extensive variables $(X_i,Y_i)$, $(i=1,2,...m)$, Eq. (\ref{dvdsdy}) can be extended as
\begin{eqnarray}
&&dY_1=\frac{\partial Y_1}{\partial X_1}d X_1+\frac{\partial Y_1}{\partial X_2}d X_2+....+\frac{\partial Y_1}{\partial X_m}d X_m, \nonumber \\
&&dY_2=\frac{\partial Y_2}{\partial X_1}d X_1+\frac{\partial Y_2}{\partial X_2}d X_2+....+\frac{\partial Y_2}{\partial X_m}d X_m, \nonumber \\
&&...\nonumber \\
&&dY_m=\frac{\partial Y_m}{\partial X_1}d X_1+\frac{\partial Y_m}{\partial X_2}d X_2+....+\frac{\partial Y_m}{\partial X_m}d X_m. \nonumber
\end{eqnarray}
If all the coefficients ${\partial Y_j}/{\partial X_i}$ $(1\leq i,j\leq m)$ take finite or infinite values on the phase boundary, we can also obtain the first or the second class of Ehrenfest equations similar to Eqs. (\ref{eeq}) or (\ref{eeq3}) for the second order phase transition, and find that the rank of the matrix $r(Z_{ic})=m-k$ ($1 \leq k<m$) for these equations can tell us that the dimensions of the phase boundary is $k$.

\begin{acknowledgments}
The authors would like to thank Liyong Ji, Qiyuan Pan and Songbai Chen for useful discussions. This work was supported by the  National Natural Science Foundation of China under Grant No. 11175065; the National Basic Research of China under Grant No. 2010CB833004; the SRFDP under Grant No. 20114306110003; Hunan Provincial Innovation Foundation For Postgraduate under Grant No CX2014xxxx; and Construct Program of the National Key Discipline.
\end{acknowledgments}

\end{document}